\documentclass{emulateapj}
\usepackage{epsfig,amsmath,amssymb,graphics,lscape}
\usepackage[colorlinks=true,  citecolor=blue, 
  linkcolor=blue,  menucolor=blue, urlcolor=blue,
  linkbordercolor={0 0 1}, frenchlinks=True, breaklinks]{hyperref}
\usepackage{multirow}

\begin{document}
\newcommand{\Ha}{H$\alpha$}
\newcommand{\Hb}{H$\beta$}
\newcommand{\Hg}{H$\gamma$}
\newcommand{\Hd}{H$\delta$}
\newcommand{\Hyd}{{\rm H}}
\newcommand{\Hae}{\Hyd\alpha}
\newcommand{\Hbe}{\Hyd\beta}
\newcommand{\Hge}{\Hyd\gamma}
\newcommand{\Hde}{\Hyd\delta}

\newcommand{\NII}{[{\rm N}\,\textsc{ii}]}
\newcommand{\OIII}{[{\rm O}\,\textsc{iii}]}
\newcommand{\OII}{[{\rm O}\,\textsc{ii}]}
\newcommand{\NeIII}{[{\rm Ne}\,\textsc{iii}]}
\newcommand{\OIIIa}{\OIII$\lambda$4363}
\newcommand{\zspecf}{z_{\rm spec}}
\newcommand{\zspec}{$\zspecf$}

\newcommand{\Pagel}{$R_{23}$}
\newcommand{\Oratio}{$O_{32}$}
\newcommand{\Te}{$T_e$}
\newcommand{\OH}{12\,+\,$\log({\rm O/H})$}
\newcommand{\OHm}{12\,+\,\log({\rm O/H})}
\newcommand{\Zsun}{$Z_{\sun}$}
\newcommand{\Msun}{$M_{\sun}$}
\newcommand{\iyr}{yr$^{-1}$}
\newcommand{\MB}{$M_B$}

\newcommand{\M}{$M_{\star}$}

\newcommand{\MZ}{\M--$Z$}
\newcommand{\MZS}{\MZ--SFR}

\newcommand{\EBV}{E(B-V)}
\newcommand{\EBVa}{$E$($B$--$V$)}

\newcommand{\Ndet}{28}
\newcommand{\SFRA}{10.7}
\newcommand{\SFRM}{4.6}
\newcommand{\SFRrange}{0.8--130 \Msun\ \iyr}

\newcommand{\MA}{$4.9\times10^{8}$}
\newcommand{\MM}{$5.0\times10^{8}$}
\newcommand{\Mrange}{$7.1\times10^7$--$2.2\times10^9$}

\newcommand{\deltaSFR}{$1.05\pm0.61$ dex} 

\newcommand{\MZoff}{$\approx$0.23 dex}

\newcommand{\doubletime}{$\approx100^{+310}_{-75}$ Myr}

\newcommand{\KSprob}{94.4\%}

\newcommand{\FMRoff}{$0.01\pm0.29$ dex}

\newcommand{\NFMRsO}{11}
\newcommand{\NFMRsT}{6}

\defcitealias{cal00}{Cal00}
\defcitealias{and13}{AM13}
\defcitealias{ly14}{Ly14}

\title{METAL-POOR, STRONGLY STAR-FORMING GALAXIES IN THE DEEP2 SURVEY: THE RELATIONSHIP
  BETWEEN STELLAR MASS, TEMPERATURE-BASED METALLICITY, AND STAR FORMATION RATE}

\shorttitle{DEEP2 Metal-Poor Galaxies}
\shortauthors{Ly et al.}

\author{Chun Ly\altaffilmark{1,4},
  Jane R. Rigby\altaffilmark{1},
  Michael Cooper\altaffilmark{2}, and
  Renbin Yan\altaffilmark{3}}

\altaffiltext{1}{Observational Cosmology Laboratory, NASA Goddard Space Flight Center,
  8800 Greenbelt Road, Greenbelt, MD 20771, USA; \email{chun.ly@nasa.gov}}
\altaffiltext{2}{Center for Galaxy Evolution, Department of Physics
  and Astronomy, UCI, Irvine, CA, USA}
\altaffiltext{3}{Department of Physics and Astronomy, University of
  Kentucky, Lexington, KY, USA}
\altaffiltext{4}{NASA Postdoctoral Fellow.}

\begin{abstract}
  \noindent
  We report on the discovery of \Ndet\ $z\approx0.8$ metal-poor galaxies in DEEP2.
  These galaxies were selected for their detection of the weak \OIIIa\ emission line,
  which provides a ``direct'' measure of the gas-phase metallicity. A primary goal
  for identifying these rare galaxies is to examine whether the fundamental
  metallicity relation (FMR) between stellar mass, gas metallicity, and star formation
  rate (SFR) holds for low stellar mass and high SFR galaxies. The FMR suggests
  that higher SFR galaxies have lower metallicity (at fixed stellar mass).
  To test this trend, we combine spectroscopic measurements of metallicity and
  dust-corrected SFRs, with stellar mass estimates from modeling the optical
  photometry. We find that these galaxies are \deltaSFR\ above the $z\sim1$ stellar
  mass--SFR relation, and $0.23\pm0.23$ dex below the local mass--metallicity
  relation. Relative to the FMR, the latter offset is reduced to 0.01 dex, but
  significant dispersion remains (0.29 dex with 0.16 dex due to measurement
  uncertainties).
  This dispersion suggests that gas accretion, star formation and chemical
  enrichment have not reached equilibrium in these galaxies.
  This is evident by their short stellar mass doubling timescale of \doubletime\
  that suggests stochastic star formation.
  Combining our sample with other $z\sim1$ metal-poor galaxies, we find a weak
  positive SFR--metallicity dependence (at fixed stellar mass) that is significant
  at 94.4\% confidence.   We interpret this positive correlation as recent star
  formation that has enriched the gas, but has not had time to drive the
  metal-enriched gas out with feedback mechanisms.
\end{abstract}



\section{INTRODUCTION}\label{sec:1}

\noindent
The chemical enrichment of galaxies, driven by star formation and modulated by gas
flows from supernova and cosmic accretion, is key for understanding galaxy formation
and evolution. The primary approach for measuring metal abundances is spectroscopy
of nebular emission lines. These emission lines can be observed in the optical and
near-infrared at $z\lesssim3$ from the ground \citep[e.g.,][]
{KK04,erb06,mai08,hai09,hay09,man09,mou11,rig11,nak12,hen13a,mom13,pir13,zah11,ly14,tro14,rey15}
and space \citep[e.g.,][]{atek11,Wel11,xia12,hen13b,whi14b}, and the
{\it James Webb Space Telescope} will extend this to $z\sim6$.

The most reliable metallicity measurements are made by measuring the flux ratio of
the \OIIIa\ line against \OIII$\lambda$5007. The technique is called the \Te\ method
because it determines the electron temperature (\Te) of the gas, and hence the
gas-phase metallicity \citep{all84}.
However, the detection of \OIIIa\ is difficult, as it is weak, almost undetectable
in metal-rich galaxies. Only 0.3\% of the Sloan Digital Sky Survey (SDSS) has
detected \OIIIa\ at signal-to-noise (S/N) $\geq2$ \citep{izo06}.

Efforts have been made to increase the number of galaxies with direct metallicities
in the local universe \citep[e.g.,][]{bro08,berg12,izo12}, and at $z\gtrsim0.2$
\citep[hereafter Ly14]{hoy05,kak07,hu09,amo14a,amo14b,ly14}; however, the total
sample size is $\sim$145 (mostly in the local universe).
  
Using the DEEP2 Galaxy Redshift Survey \citep{dav03,new13}, we are conducting
studies that utilize \OIIIa\ detections in individual galaxies and from stacked
spectra. In this paper, we focus on first results from individual \OIIIa\
detections in \Ndet\ galaxies. Our sample of \Ndet\ galaxies substantially
increases the number of $z\geq0.25$ galaxies with S/N$\geq$3 \OIIIa\ detections
by 44\% (64 to 92).

While the selection function is biased toward metal-poor galaxies, this
galaxy population is of significant interest for understanding galaxy
evolution. Their low metallicity suggests that they are either (1) in their
earliest stages of formation, (2) accreting metal-poor gas, or (3) undergoing
significant metal-enriched gas outflows. Metal-poor galaxies have been poorly
studied due to the difficulty of identifying them. Thus, the majority of
previous high-$z$ metallicity studies have only used strong-line metallicity
calibrations. These calibrations are problematic for high-$z$ galaxies due to
suspected differences in the physical conditions of the interstellar gas
\citep[e.g.,][]{liu08,kew13a,kew13b}, but see also \cite{jun14} for a different
interpretation. These differences, if present, may be incorrectly interpreted
as evolution in the metal content. Thus, obtaining \Te-based metallicities at
high redshifts is the next logical step, and these \OIIIa\ detections can
potentially be used to recalibrate the strong-line diagnostics.

While the \Te\ method is affected by properties of the ionized gas
\citep[e.g., optical depth, density, ionization parameter, non-equilibrium
  electron energy;][]{nic14}, such effects also apply to strong-line
diagnostics as \cite{nic14} demonstrated. Thus, while the \Te\ method is
less ``direct'' than was initially determined \citep{sea54}, measuring the
electron temperature currently remains the preferred way to determine gas
metallicities.

In this paper, we focus on the relationship between stellar mass,
gas-phase metallicity, and SFR for our sample of \Ndet\ metal-poor galaxies.
This relationship has received significant interest as \cite{ell08} found
that at a given stellar mass, lower-metallicity galaxies in the local
universe tend to have higher SFRs. Thus, while the stellar mass--metallicity
relation is tight \citep[$\sim$0.1 dex;][]{tre04}, it may be a projection of
a non-evolving three-dimensional relationship between stellar mass (\M),
gas-phase metallicity ($Z$), and SFR \citep[e.g.,][]{lar10,man10,hunt12}.
  
However, the existence of a \MZS\ relation (dubbed the
``fundamental metallicity relation'' or ``FMR'') remains controversial, as
recent studies have yielded results that agree or disagree with predictions
(see \citealt{sal14} and \citealt{rey15} for a review).
Moreover, the \MZS\ relation has yet to be tested with large samples of
metal-poor ($Z\lesssim0.25$ \Zsun) galaxies, especially at higher redshift.
The largest high-$z$ metal-poor sample is that of \citetalias{ly14} from the
Subaru Deep Field \citep[SDF;][]{kas04,ly07}, which detected \OIIIa\ in 20
galaxies at $z\sim0.4$--1. In this study, they found evidence that galaxies
with the highest specific SFR (SFR/\M, hereafter sSFR) were not necessary more
metal-poor. This result, based on 20 galaxies, requires further confirmation
with our DEEP2 sample.

In this study, we compare our work with \cite{and13} (hereafter AM13)
who stacked $0.027<z<0.25$ SDSS spectra in bins of SFR and \M\ to
obtain average \OIIIa\ measurements. We note that the primary selection
functions of \citetalias{and13} (magnitude-limited) and our study
(emission-line strengths) are different. In a forthcoming paper, we will
follow the same approach as \citetalias{and13} of stacking the spectra of
a few thousand $z\sim$0.8 DEEP2 galaxies, allowing for a more self-consistent
comparison. For now, we investigate the \MZS\ relation with individual
\Te\ metallicity measurements.
Throughout this Letter, we adopt a cosmology with $\Omega_{\Lambda}=0.7$,
$\Omega_M=0.3$, and $h=0.7$, a \cite{cha03} initial mass function (IMF), and
a solar metallicity of \OH\ = 8.69. 


\begin{figure*}
  \epsscale{1.1}
  \plotone{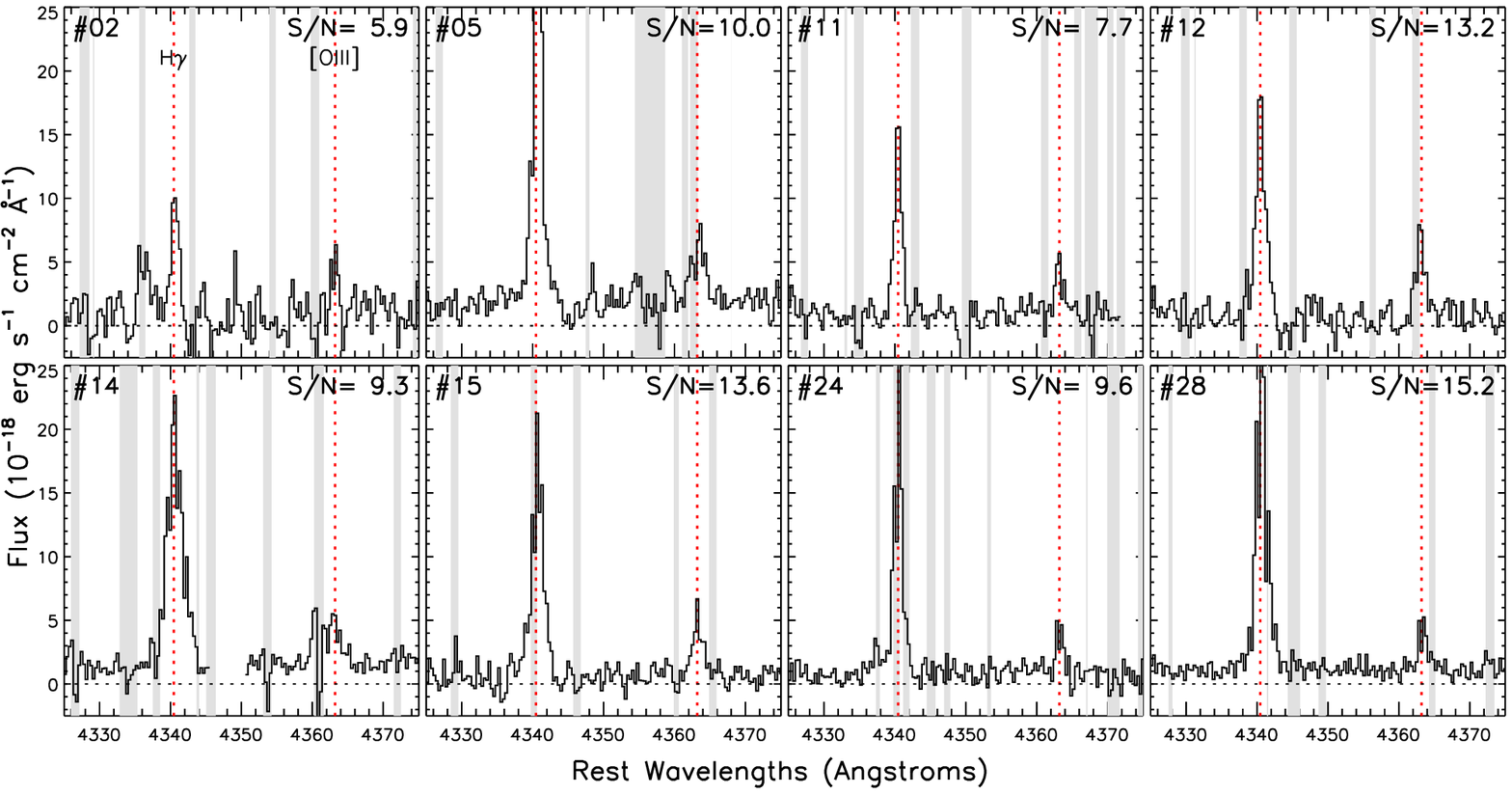}
  \vspace{-0.15cm}
  \caption{Detections of \OIIIa\ in $z\sim0.8$ DEEP2 galaxies. The Keck/DEIMOS
    spectra for 8 of \Ndet\ galaxies are shown by the solid black lines, with
    vertical red dashed lines indicating the locations of \Hg$\lambda$4340 and
    \OIIIa. OH skylines are indicated by the grey shaded regions. The
    signal-to-noise of \OIIIa\ detections is reported in the top right.}
  \label{fig:spec}
\end{figure*}

\section{THE SAMPLE}
\label{sec:deep2}

\noindent
The DEEP2 Survey has surveyed $\sim$3 deg$^2$ over four fields using the DEIMOS
multi-object spectrograph \citep{fab03} on the Keck-II telescope. The survey
has provided optical ($\approx$6500--9000\AA) spectra for $\sim$53,000 galaxies
brighter than $R_{\rm AB}=24.1$, and precise redshifts for $\sim$70\% of targeted
galaxies. An overview of the survey can be found in \cite{new13}.

Using the fourth data release (DR4),\footnote{\url{http://deep.ps.uci.edu/dr4/home.html}.}
we select 37,396 sources with reliable redshifts (quality flag $\geq$3). We
consider those with spectral coverage that spans 3720--5010 \AA\ (rest-frame).
This enable us to determine metallicity from oxygen and hydrogen emission lines
(\OII$\lambda\lambda$3726,3729, \OIII$\lambda\lambda\lambda$4363,4959,5007, and
\Hb), and further limits the sample to 4,140 galaxies at $z=0.697$--0.859
(average: 0.779).

\begin{figure*}
  \epsscale{1.1}
  \plotone{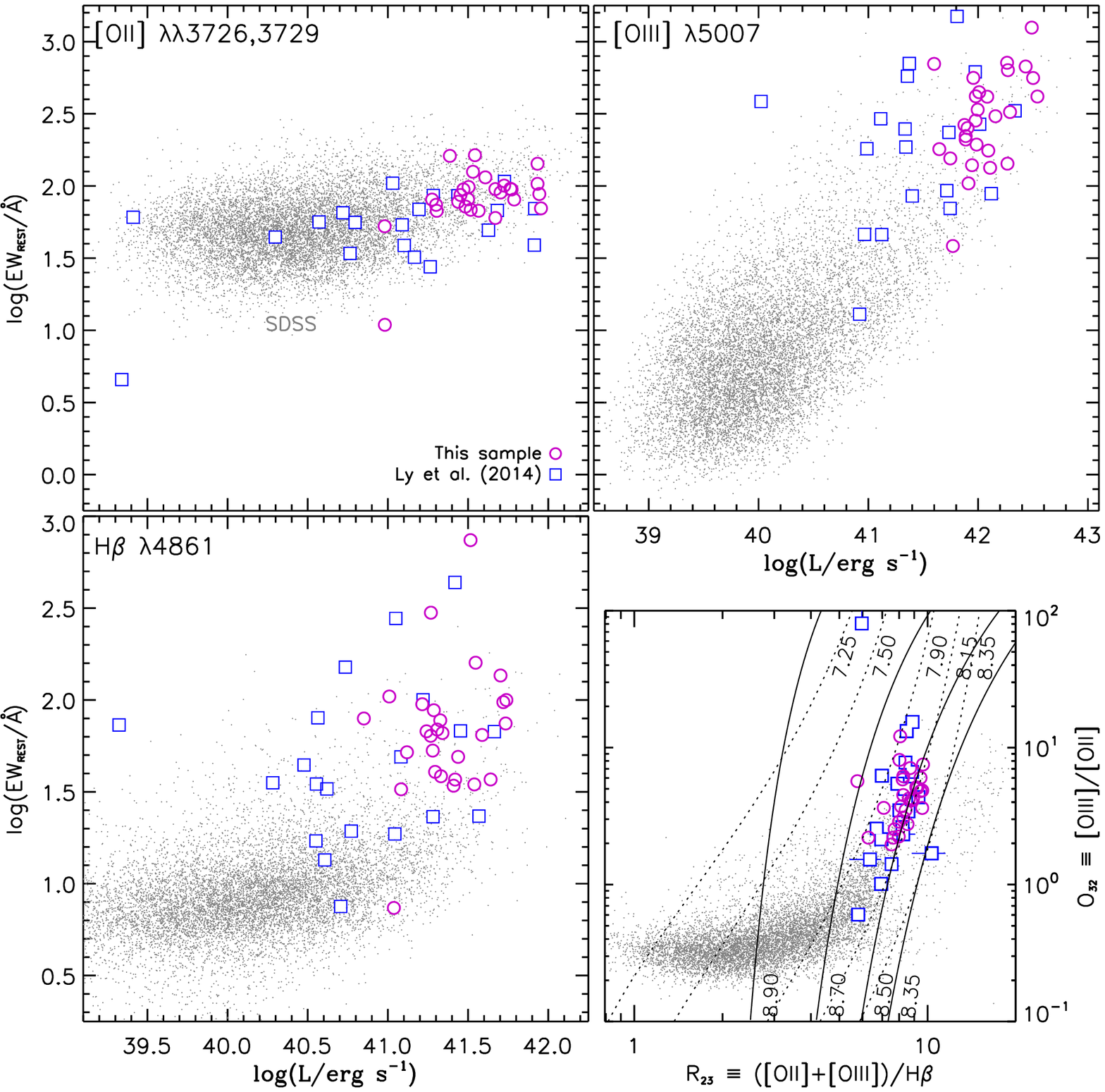}
  \vspace{-0.25cm}
  \caption{Emission-line luminosities, flux ratios, and rest-frame EWs for our
    \OIIIa\ sample (purple circles).  All luminosities and flux ratios are
    observed, before correction for dust attenuation. Gray points illustrate
    the SDSS DR7 emission-line sample. The lower right panel shows the
    metallicity-sensitive (\Pagel) and ionization parameter-sensitive (\Oratio)
    emission-line ratios. Photoionization models from \cite{m91} are overlaid
    for metallicities between \OH\ = 7.25 and \OH\ = 8.9. Solid (dotted) curves
    are for metallicities on the ``upper'' (``lower'') \Pagel\ branch. Overlaid
    as blue squares is the \OIIIa-detected sample from \citetalias{ly14}.}
  \label{fig:lines}
\end{figure*}

We follow the approach of \citetalias{ly14} that fits emission lines with Gaussian
profiles using the IDL routine \textsc{mpfit} \citep{mpfit}. Spectroscopic redshifts
are used as priors for the location of emission lines. With measurements of
emission-line fluxes and the noise in the spectra (measured from a 200 \AA\ region
around each line), we select those with \OIIIa\ and \OIII$\lambda$5007 detected at
S/N$\geq$3 and S/N$\geq$100, respectively. 
This yields an initial sample of 54 galaxies. We inspect each spectrum and remove
26 galaxies from our sample, primarily because of contamination from OH sky-lines.
This leaves \Ndet\ galaxies. 
One source (\#21) was observed twice. The other spectrum also detected \OIIIa\
at lower S/N, so the better spectrum is used in our analysis.
Compared to the previous DEEP2 sample \citep{hoy05}, we confirmed two, thus 26
galaxies in our sample are newly identified.
Detections of \OIIIa\ are shown in Figure~\ref{fig:spec}, and galaxy properties
are provided in Table \ref{tab:1}. We illustrate in Figure~\ref{fig:lines} the
emission-line luminosities, rest-frame equivalent widths (EWs), and \Oratio\
$\equiv$ \OIII/\OII\ and \Pagel\ $\equiv$ (\OII+\OIII)/\Hb\ flux ratios
\citep{pag79,m91}, and compare our sample to local galaxies and other
\OIIIa-detected galaxies \citepalias{ly14}. 

\subsection{Flux Calibration}

The publicly released data of DEEP2 are not flux-calibrated, which is problematic
for measuring the 4363-to-5007 ratio, and hence \Te. To address this limitation,
we use proprietary IDL codes developed by Jeffrey Newman, Adam Walker, and
Renbin Yan of the DEEP2 team. These codes take into account the overall
throughput, quantum efficiency of the eight CCD detectors, apply coarse telluric
corrections for atmospheric absorption bands, and use the $R$ and $I$ DEEP2
photometry to transform the spectrum to energy units. The DEEP2 team has
demonstrated that the calibration is reliable at the 10\% level when compared to
SDSS stars observed by DEEP2.


\section{DERIVED PROPERTIES}
\label{sec:prop}

\subsection{Dust Attenuation Correction from Balmer Decrements}
\label{sec:dust}

\begin{figure*}
  \epsscale{1.1}
  \plotone{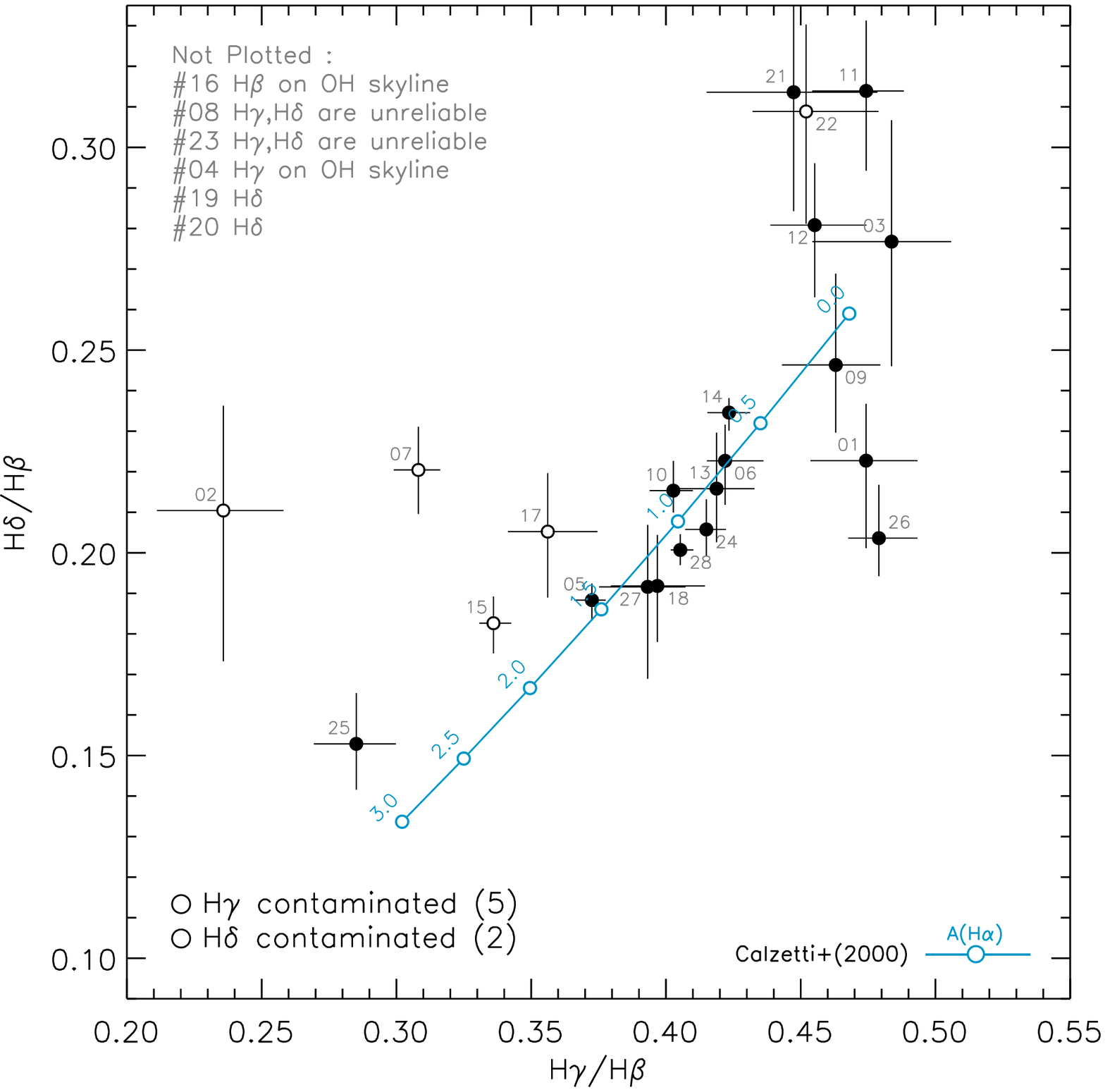}
  \vspace{-0.25cm}
  \caption{Balmer decrements (\Hg/\Hb\ and \Hd/\Hb) for our \OIIIa\ sample.
    Reliable measurements are shown by the filled circles, while those
    affected by contamination from OH sky-line emission are shown as open
    circles. Blue circles and curve show the effects on the Balmer
    decrements with increasing dust reddening following \citetalias{cal00}.
    Values adjacent to blue circles indicate $A$(\Ha).
    The significant scatter in the upper right is due to less reliable
    \Hd/\Hb\ measurements. These galaxies all have \Hg/\Hb\ measurements
    that are consistent with $A$(\Ha) $\sim0$.}
  \label{fig:dust}
\end{figure*}

\noindent
To correct the emission-line fluxes for dust attenuation, we use Balmer
decrement measurements. At $z\sim0.8$, the existing DEEP2 optical spectra
measure \Hb, \Hg, and \Hd.
While these lines are intrinsically weak compared to \Ha,\footnote{\Ha\ is
redshifted beyond the optical spectral coverage.} our galaxies possess high
emission-line EWs, which result in 22, 26, and \Ndet\ galaxies having \Hd,
\Hg, and \Hb\ detected at S/N$\geq$10, respectively. The significant
detections enable dust attenuation measurements of
$\sigma$(A(\Ha))$\,\approx0.1$ mag (average from \Hg/\Hb).

A problem encountered with Balmer emission lines is the underlying stellar
absorption. Our examination of each spectrum reveal weak stellar absorption,
making it difficult to obtain reliable fits to the broad wings of absorption
lines. To address this limitation, we stack our spectra. Here, the continuum
(around each Balmer line) is normalized to one, and an average is computed
with the exclusion of spectral regions affected by OH sky-line emission.
Stellar absorption is detected in \Hd, and is consistent with an EW$_{\rm rest}$
correction of 1 \AA. For our entire sample, we adopt an EW$_{\rm rest}$
correction of 1 \AA\ for \Hb, \Hg, and \Hd. With these corrections for
stellar absorption, we illustrate the Balmer decrements in
Figure~\ref{fig:dust}.

Assuming that the hydrogen emission originates from an optically thick
\textsc{H ii} region obeying Case B recombination, the intrinsic Balmer
flux ratios are (\Hg/\Hb)$_0$ = 0.468 and (\Hd/\Hb)$_0$ = 0.259.
Dust absorption alters these observed ratios as follows:
\begin{equation}
  \frac{(\Hyd n/\Hbe)_{\rm obs}}{(\Hyd n/\Hbe)_0} = 10^{-0.4\EBV[k(\Hyd n)-k(\Hbe)]},
  \label{eqn:balmer}
\end{equation}
where \EBVa\ is the nebular color excess, and $k(\lambda)\equiv A(\lambda)$/\EBVa\
is the dust reddening curve. We illustrate in Figure \ref{fig:dust} the
observed Balmer decrements under the \cite{cal00} (hereafter Cal00) dust
reddening formalism.
We find that our Balmer decrements are consistent with \citetalias{cal00}.
For the remainder of our Letter, all dust-corrected measurements
adopt \citetalias{cal00} reddening.

Our color excesses, are determined mostly (20/\Ndet) from \Hg/\Hb. For five
galaxies, we use \Hd/\Hb\ since \Hg\ suffers from contamination from OH
skylines.
For the remaining 3 galaxies, the dust reddening could not be determined from
either Balmer decrement (they were both affected by OH sky-line emission).
For these galaxies, we assume $\EBV=0.22\pm0.23$ mag
($A(\Hae)\approx0.73\pm0.75$ mag), which is the average of our sample, and is
consistent with other emission-line galaxy samples
\citep[e.g.,][]{ly12a,dom13,mom13}.
For Balmer decrements that imply negative reddening (6 cases), we adopt $\EBV=0$
with measurement uncertainties based on Balmer decrement uncertainties.

\subsection{\Te-based Metallicity Determinations}

\noindent To determine the gas-phase metallicity for our galaxies, we follow
previous direct metallicity studies and use the empirical relations of
\cite{izo06}. Here, we briefly summarize the approach, and refer readers
to \citetalias{ly14} for more details.
First, the O$^{++}$ electron temperature, \Te(\OIII), can be estimated using
the nebular-to-auroral \OIII\ ratio,
\OIII\,$\lambda\lambda$4959,5007/\OIII\,$\lambda$4363.
We correct the above flux ratio for dust attenuation (Section~\ref{sec:dust}).
We also apply a 5\% correction, since \Te\ determinations from \cite{izo06} are
found to be overestimated \citep{nic13}.

Our \OIII\ measurements have a
very large dynamic range. The strongest (weakest) \OIIIa\ line is 6.5\%
(0.7\%) of the \OIII\,$\lambda$5007 flux. We find that the average (median)
$\lambda$4363/$\lambda$5007 flux ratio for our sample is 0.018 (0.015).
The derived \Te\ for our galaxy sample spans (1--3.1)$\times10^4$ K.

To determine the ionic abundances of oxygen, we use two emission-line flux
ratios, \OII\,$\lambda\lambda$3726,3729/\Hb\ and
\OIII\,$\lambda\lambda$4959,5007/\Hb.
For our metallicity estimation, we adopt a standard two-zone temperature
model with \Te(\OII) = 0.7\Te(\OIII) + 3000 \citepalias{and13}, to enable
direct comparisons to local studies.
In computing O$^+$/H$^+$, we also correct the \OII/\Hb\ ratio for dust
attenuation. We do not correct \OIII/\Hb\ since the effects are negligible.

Since the most abundant ions of oxygen in \textsc{H ii} regions are O$^+$
and O$^{++}$, the oxygen abundances are given by
${\rm O/H}=({\rm O}^++{\rm O}^{++})/{\rm H}^+$.
In Table \ref{tab:1}, we provide estimates of \Te(\OIII), and de-reddened
metallicity for our sample. Our most metal-poor systems are \#04 and \#08,
and can be classified as extremely metal-poor galaxies ($\leq$0.1 \Zsun).

\subsection{Dust-Corrected Star Formation Rates}

\noindent
In addition to gas-phase metallicity, our data allow us to determine
dust-corrected SFRs using the hydrogen recombination lines, which are
sensitive to the shortest timescale of star formation, $\lesssim$10 Myr.

Assuming a \cite{cha03} IMF with masses of 0.1--100 \Msun, and solar
metallicity, the SFR can be determined from the observed \Hb\ luminosity
\citep{ken98}:
\begin{equation}
  \frac{{\rm SFR}}{M_{\sun}~{\rm yr}^{-1}} = 4.4\times10^{-42} \times 2.86 \times 10^{0.4 A(\Hbe)}
  \frac{L(\Hbe)}{{\rm erg~s}^{-1}},
\end{equation}
where $A(\Hbe)=4.6$\EBVa. This relation overestimates the SFR at low
metallicities due to the dependence of a stronger ionizing radiation
field on lower metallicity. Since our galaxies have $Z\approx0.2Z_{\sun}$,
we reduce the SFRs by 37\% \citep{hen13b}.
Our SFR estimates are summarized in Table \ref{tab:1} and are illustrated
in Figure~\ref{fig:5}. We find that our galaxies have dust-corrected
SFRs of \SFRrange\ with an average (median) of \SFRA\ (\SFRM) \Msun\ \iyr.

\subsection{Stellar Masses from SED Modeling}

\noindent
To determine stellar masses, we follow the common approach of modeling the
spectral energy distribution (SED) with stellar synthesis models
\citep[e.g.,][]{sal07,ly11,ly12b}.
The eight-band photometric data include $BRI$ imaging from the
Canada-France-Hawaii Telescope (CFHT) for the DEEP2 survey \citep{coil04}.
In addition, publicly available $ugriz$ imaging from the CFHT Legacy Survey
is available in Field \#1 (Extended Groth Strip), and Fields \#3--4 are
located in the SDSS deep survey strip (Stripe 82) for
$u\arcmin g\arcmin r\arcmin i\arcmin z\arcmin$ imaging. Thus, 18 of 27
galaxies have eight optical imaging bands. These photometric data that we
use have been compiled by \cite{mat13}.
While our galaxies have low stellar masses (as demonstrated below), the
imaging data are fairly deep. The $R$-band imaging, for example, reaches
an 8$\sigma$ AB limit of 24.5 \citep{new13}, and our galaxies are on
average 1.2 AB mag brighter than this limit.
  
To extend the wavelength coverage, we cross-matched our sample against
the catalog of \cite{bun06}, which contains $JK$ photometry. Unfortunately,
only two galaxies have a match. This is not a surprise since many of our
galaxies have low stellar masses. While photometric data redward of 5500\AA\
are unavailable, \cite{zah11} demonstrated that stellar mass estimates from
$BRI$ photometry are consistent with those obtained from $BRIK_s$, suggesting
that the existing optical data are sufficient for stellar mass estimates.
We also note that the lack of $ugriz$ imaging does not significantly hamper out
SED modeling for 9 galaxies, as we compared stellar mass estimates derived from
$BRI$-only and $BRI$+$ugriz$ photometry for two-thirds of our sample, and find
consistent results.
Future efforts will include acquiring {\it Spitzer} infrared data to provide
more robust stellar mass estimates.

\begin{figure}
  \epsscale{1.1}
  \plotone{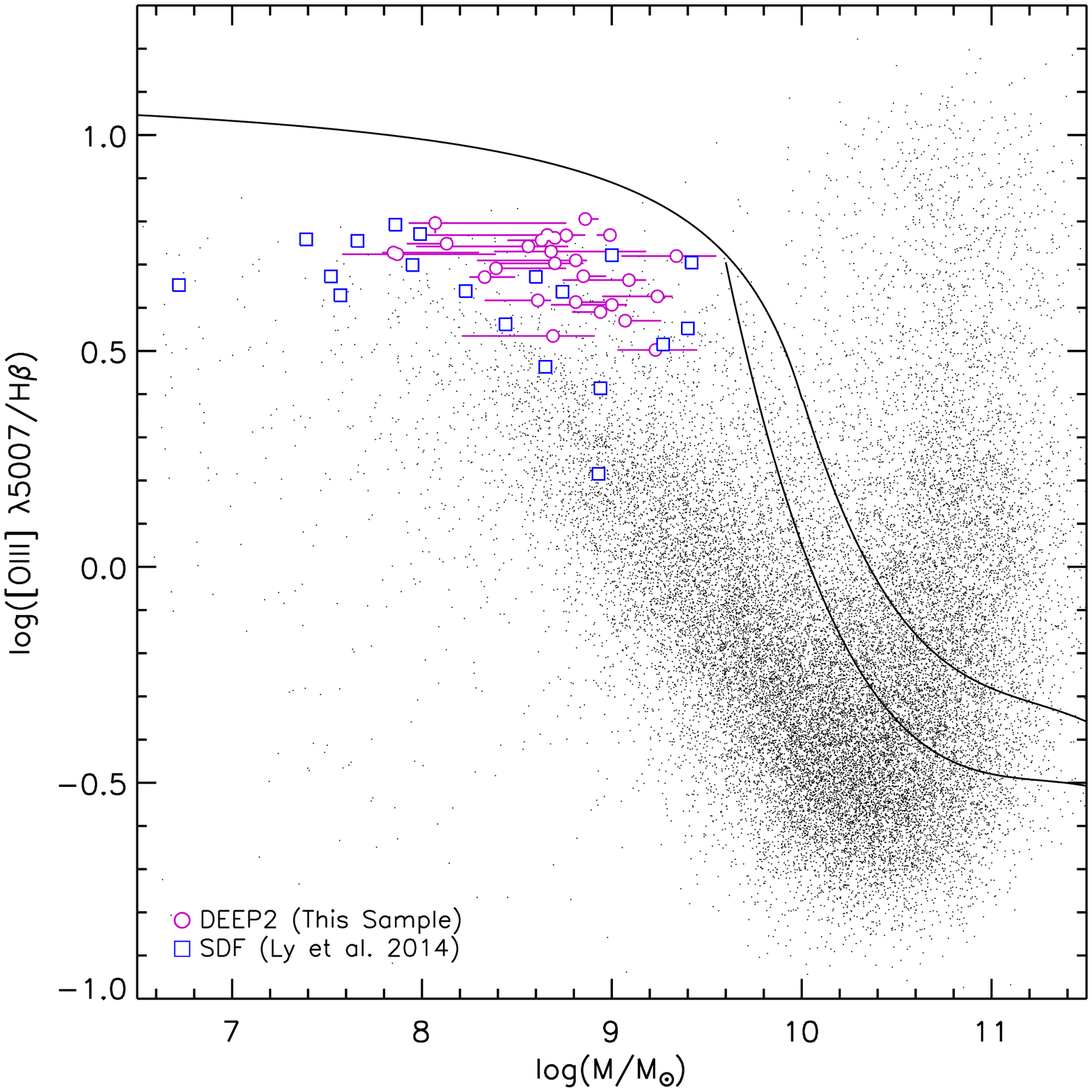}
  \vspace{-0.2cm}
  \caption{\OIII/\Hb\ flux ratio (``excitation'') as a function of stellar mass
    \citep[i.e., the MEx diagram;][]{jun14}. The DEEP2 \OIIIa\ sample is shown
    by the purple circles. Overlaid as blue squares is the \citetalias{ly14}
    \OIIIa\ sample from the SDF. The SDSS DR7 sample is illustrated in black.}
  \label{fig:4}
\end{figure}

\begin{figure*}
  \epsscale{1.0}
  \plotone{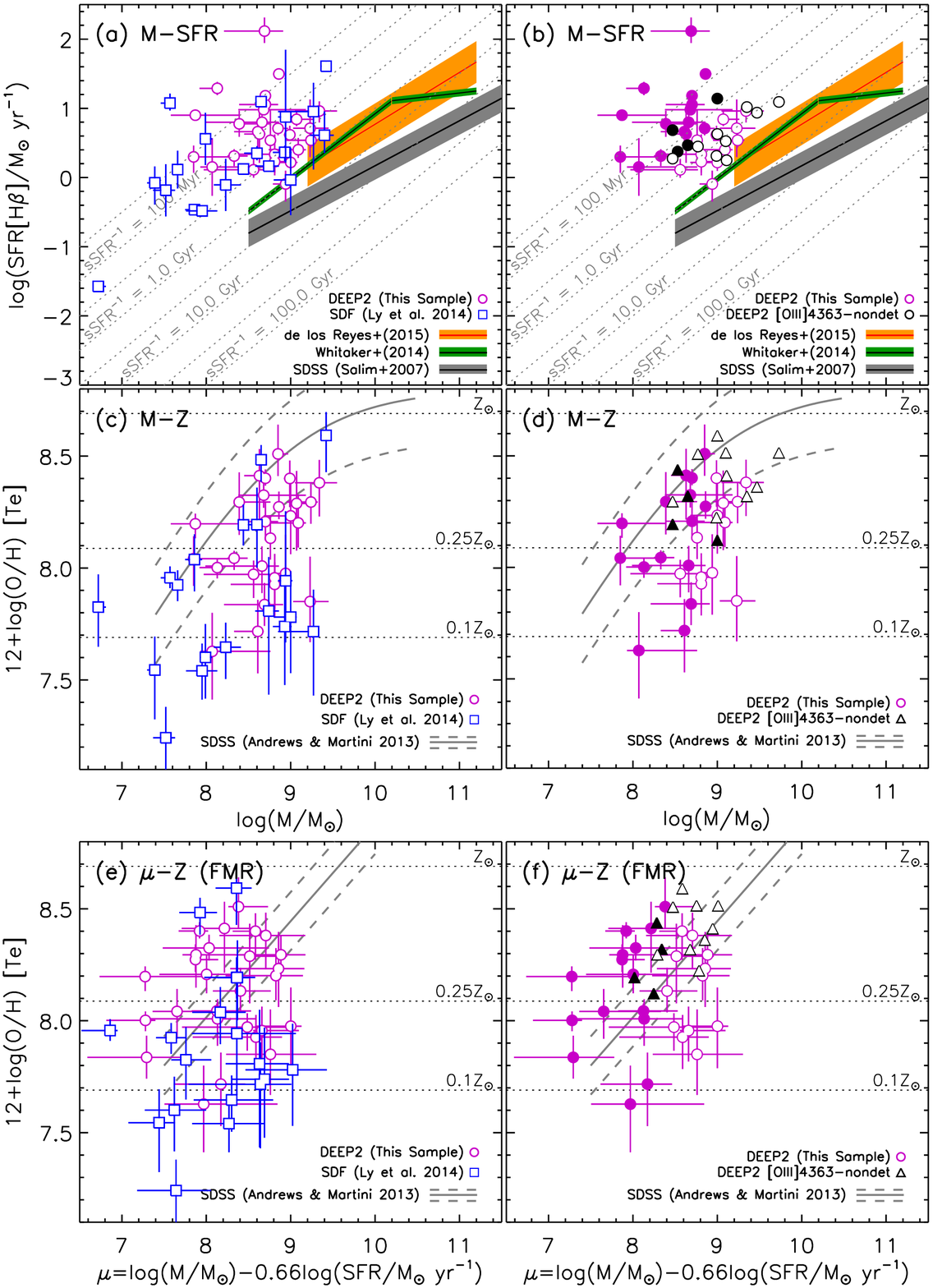}
  \caption{Relations between stellar mass and (top) dust-corrected \Hb\ SFR,
    (middle) metallicity, and (bottom) metallicity and dust-corrected \Hb\ SFR
    (i.e., the FMR). The DEEP2 \OIIIa\ sample is shown by the purple circles.
    Overlaid as blue squares on the left panels is the \citetalias{ly14} \OIIIa\
    sample from the SDF. Results from the SDSS sample (\citealt{sal07};
    \citetalias{and13}) is illustrated in gray. For comparisons, we also overlay
    in the top panels the stellar mass--SFR relation of \Ha-selected galaxies at
    $z=0.8$ \citep{rey15} and mass-selected star-forming galaxies at $z=0.5$--1
    \citep{whi14} in orange and green, respectively. For direct metallicity
    comparisons, we illustrate the results of \citetalias{and13}, which stacked
    spectra to measure average \Te-based metallicities. DEEP2 galaxies with
    reliable non-detections of \OIIIa\ are shown as black circles and triangles
    (lower limit on metallicity). The DEEP2 samples are separated by low and
    high SFRs (see text) as open and filled symbols, respectively.}
  \label{fig:5}
\end{figure*}

To model the SED, we use the Fitting and Assessment of Synthetic Templates code
\citep{kri09} with \cite{bc03} models and adopt a \cite{cha03} IMF,
exponentially-declining star formation histories (SFHs; i.e., $\tau$ models),
one-fifth solar metallicity, and \citetalias{cal00} reddening.
We also correct the broad-band photometry for the contribution of nebular
emission lines following the approach described in \citetalias{ly14}. This
correction reduces the stellar mass estimates by 0.2 dex (average).
To determined stellar mass uncertainties, we conduct Monte Carlo realizations
within FAST. Here, data points are randomly perturbed 100 times (based on the
photometric uncertainties) and the SEDs are re-fitted, yielding a probability
distribution function for stellar mass.
The stellar masses are provided in Table \ref{tab:1} and are illustrated in
Figures~\ref{fig:4}--\ref{fig:5}. The average (median) stellar masses are \MA\
\Msun\ (\MM\ \Msun) and span \Mrange\ \Msun.


\section{RESULTS}
\label{sec:results}

\subsection{Excitation Properties}
\noindent
Figure~\ref{fig:4} illustrates the \OIII$\lambda$5007/\Hb\ flux ratios and
stellar masses along the ``Mass--Excitation'' \citep[MEx;][]{jun14} diagram.
The MEx is used as a substitute for the \cite{bal81} diagnostic
diagram when \NII$\lambda$6583/\Ha\ is unavailable.
It can be seen that these galaxies have high \OIII/\Hb\ ratios, $5.0\pm0.9$.
All of them are classified as star-forming galaxies by falling below the
solid black line.
Compared to other metal-poor galaxies \citepalias[blue squares]{ly14},
the DEEP2 galaxies have similar excitation properties, but are $\approx$0.4
dex more massive.
Compared to UV- and mass-selected $z\sim2$ galaxies \citep[e.g.,][]{sha14,ste14},
our measured \OIII/\Hb\ ratios are higher by a factor of 1.25--2.5.
Their strong-line oxygen ratios, $R_{23}$ and $O_{32}$, are consistent with
$z\sim2$ galaxies from \cite{sha14}. 

\subsection{Relationship between Mass, Metallicity, and SFR}
\noindent
Figure~\ref{fig:5}(a) compares the dust-corrected instantaneous SFRs against
the stellar mass estimates. Here we compare our work against mass-selected
galaxies at $z\sim1$ \citep{whi14b} and \Ha-selected galaxies at $z\approx0.8$
\citep{rey15}.
Our galaxies are located \deltaSFR\ above these \M--SFR relations
with SFR/\M\ of 10$^{-8.0\pm0.6}$ yr$^{-1}$.
This significant SFR offset is also seen for metal-poor galaxies
from \citetalias{ly14}.
By requiring \OIIIa\ detections, both \OIIIa\ studies are
biased toward high-EW emission lines (see Figure~\ref{fig:lines}),
which correspond to a higher sSFRs.

Figure~\ref{fig:5}(c) illustrates the \MZ\ relation. Here we compare our
results against \citetalias{and13}.
It demonstrates that while a subset of our galaxies is consistent with
\citetalias{and13}, a significant fraction (60\%) are located below
the relation at more than 0.22 dex
\citepalias[1$\sigma$;][]{and13}, by as much as --0.76 dex. This
results in an average $Z$ offset for the sample of --$0.28\pm0.23$ dex.
Our \MZ\ relation result is consistent with \citetalias{ly14} (blue squares),
who also found that half of their sample falls below the local \MZ\ relation.

The FMR was introduced to describe the dependence between \M,
$Z$, and SFR in local galaxies, and was extended to explain higher redshift
galaxies. \cite{man10} was one of the first studies to parameterize this
dependence by considering a combination of stellar mass and SFR:
\begin{equation}
  \mu = \log{(M_{\star})} - \alpha\log{(\rm SFR)},
\end{equation}
where $\alpha$ is the coefficient that minimizes the scatter against metallicity.
Figure~\ref{fig:5}(e) illustrates the $\mu$ projection of the \MZS\ relation 
with $\alpha=0.66$ \citepalias{and13}.
It can be seen that our sample is consistent (\FMRoff) with the local FMR;
however significant dispersion remains. The dispersion is greater than our
\MZ\ comparison and the average measurement uncertainties of $\approx$0.16
dex with respect to the FMR.

The local \MZS\ relation suggests that higher SFR galaxies have lower metallicity
at fixed stellar mass. To examine if this is correct, we split our sample by
high and low sSFRs, and perform Kolmogorov-Smirnoff (K-S) tests to determine if
these two distributions are different. These two samples are illustrated in panels
(b), (d) and (f) in Figure~\ref{fig:5} as filled (high-SFR) and open (low-SFR)
symbols.
The sample is divided at the median $\Delta[\log({\rm SFR})]$, which is the
amount of deviation relative to the \cite{whi14} \M--SFR relation.
This relative SFR offset follows the non-parametric approach of \cite{sal14}.

One concern with conducting a K-S test is the selection bias of requiring the
detection of \OIIIa. More specifically, detection of this line primarily
depends on the electron temperature (or gas metallicity), which corresponds
to the $\lambda$5007/$\lambda$4363 line ratio, and the dust-corrected SFR, which
determines the overall normalization of the emission-line strengths. At high
SFRs, the probability of detecting \OIIIa\ is greater for a wide range of
metallicity. This range of metallicity reduces toward lower metallicity such
that only metal-poor galaxies with low SFRs can be detected in an emission-line
flux limited survey.
  
\begin{figure}
  \epsscale{1.1}
  \plotone{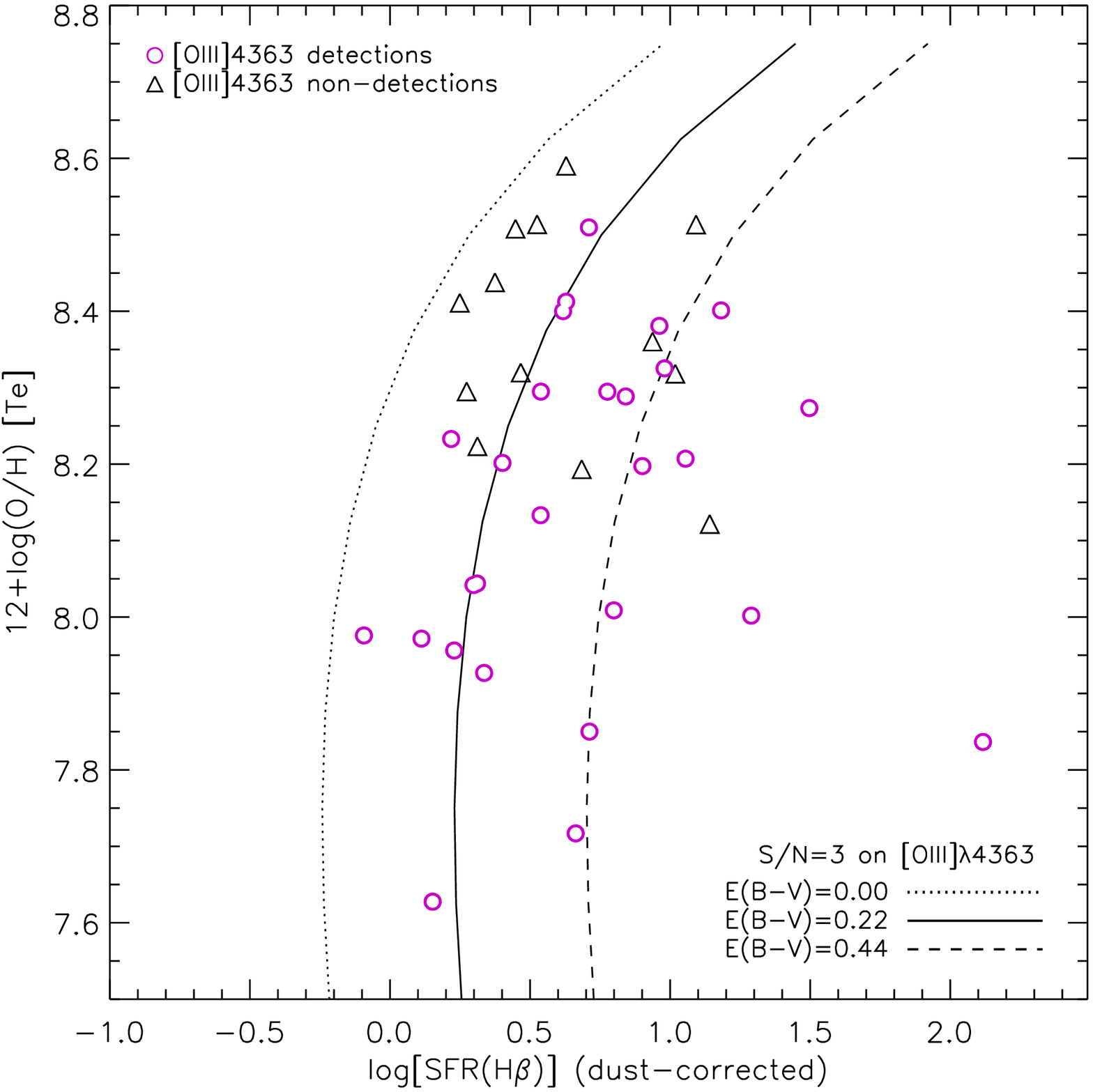}
  \caption{Gas-phase metallicity as a function of dust-corrected SFR for
    our \OIIIa\ sample (purple circles) and a sample of reliable
    \OIIIa\ non-detections at S/N$<$3 (black triangles). The triangles
    are lower limits on metallicity. The dotted, solid, and
    dashed lines correspond to a S/N=3 limit on \OIIIa\ of
    $3.4\times10^{-18}$ erg s$^{-1}$ cm$^{-2}$ for three dust extinction
    possibilities, \EBV\ = 0.0, 0.22, and 0.44, respectively. These
    extinction values span the dispersion seen in our sample.}
  \label{fig:6}
\end{figure}

This selection bias is demonstrated in Figure~\ref{fig:6}, which illustrates
the gas-phase metallicity as a function of dust-corrected SFR. The black curves
correspond to the average \OIIIa\ S/N=3 detection limit of $3.4\times10^{-18}$
erg s$^{-1}$ cm$^{-2}$. This limit was determined from measuring the rms in
4,140 spectra in areas adjacent to where \OIIIa\ is expected to be detected.
In determining the metallicity--SFR dependence, we consider (1) an ionic oxygen
abundance ratio (O$^{++}$/O$^+$) of unity, which is the average for our DEEP2
\OIIIa\ sample, and (2) a temperature--metallicity relation of \OH\ = 9.51 -
1.03(\Te/10$^4$ K). The latter is empirically determined from our DEEP2 \OIIIa\
galaxies.

To account for the $Z$--SFR selection bias, we include reliable \OIIIa\
non-detections within our K-S analyses. First, we consider all DEEP2 galaxies
with S/N$\geq$100 on \OIII$\lambda$5007 and a non-detection (S/N$<$3) on \OIIIa.
This sample of 126 galaxies is then vetted for unreliable limits because
the \OIIIa\ emission line either falls on an OH skyline, the atmospheric A-band,
or a CCD gap. This limited the sample to 79 galaxies. While the above
\OIII$\lambda$5007 S/N cut is strict, for many of these galaxies, a strong lower
limit on the metallicity is not available. This is because a S/N=100 corresponds
roughly to 0.03 on $\lambda$4363/$\lambda$5007 or \Te\ $\approx2\times10^4$ K.
Thus, we further restrict our non-detection sample to 13 galaxies with
\OIII$\lambda$5007 S/N$\geq$200. These galaxies are overlaid in Figure~\ref{fig:6}
and panels (b), (d), and (f) in Figure~\ref{fig:5} as either black circles or
black triangles (S/N=3 lower limit on metallicity). It can be seen that the
majority of these galaxies are located to the left of the S/N=3 line for
\EBV\ = 0.22 mag in Figure~\ref{fig:6}. Also, Figure~\ref{fig:5}
illustrates that these galaxies have higher stellar masses (0.3 dex) than our
detected sample, and lower SFRs at a given stellar mass. As expected, these
galaxies have higher metallicity with lower limits above \OH\ $\approx8.2$
(average: 8.37).

\begin{figure}
  \epsscale{1.1}
  \plotone{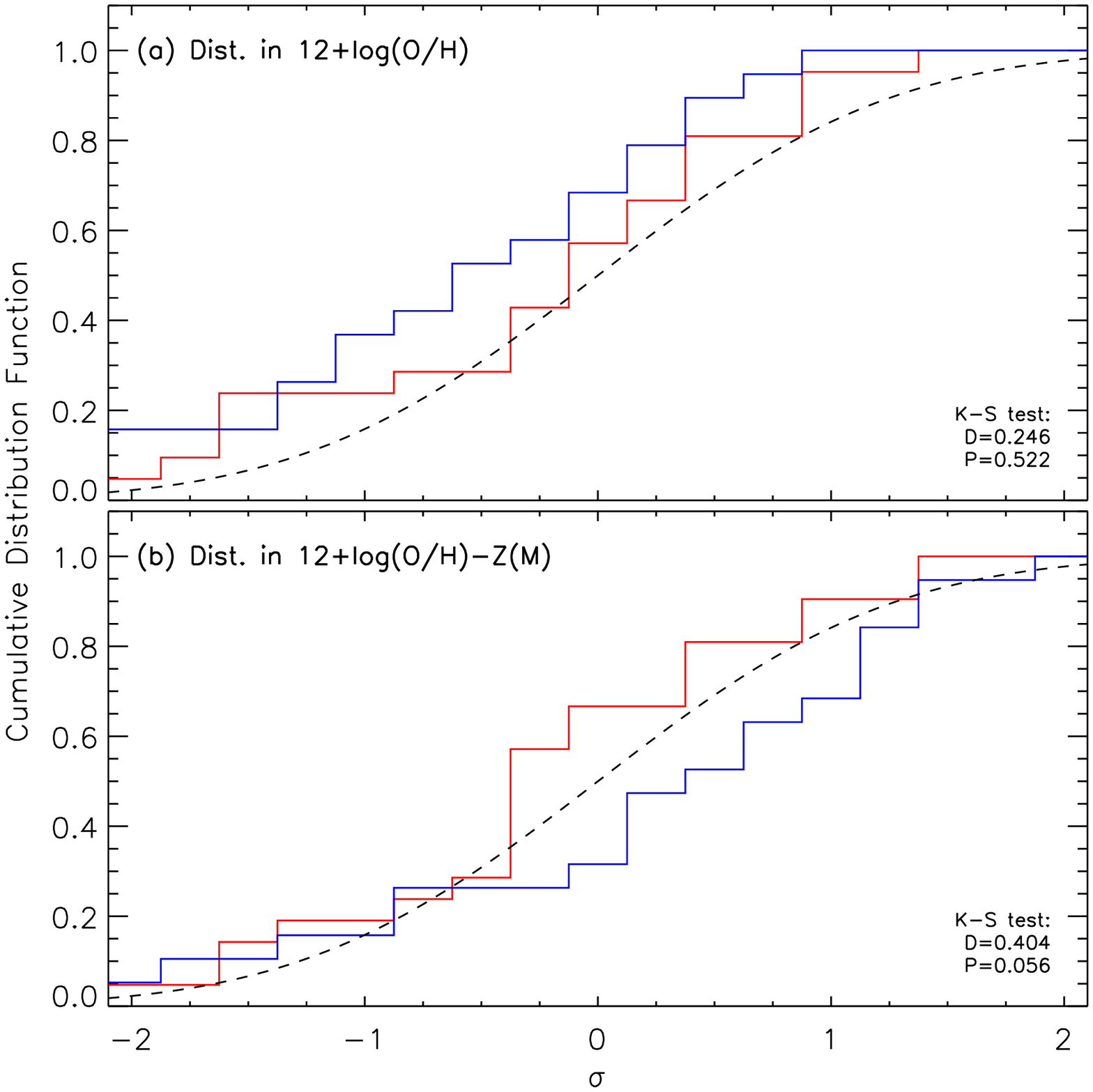}
  \caption{Cumulative distribution functions in (a) metallicity and
    (b) metallicity relative to the \citetalias{and13} \MZ\ relation, for two
    samples with low (red line) and high SFRs (blue line). The K-S statistics
    (D) and the probability that the two distributions are identical (P) are
    given in the lower right-hand corner. Panel (b) shows that the two
    distributions are different at \KSprob\ confidence.}
    \label{fig:7}
\end{figure}

For our K-S tests, we compare the $\log({\rm O/H})$ distributions for the
low- and high-SFR samples, finding that these two distributions are similar
with the lower SFR galaxies having slightly higher metallicity (see
Figure~\ref{fig:7}(a)). However, as Figure~\ref{fig:5}(b) shows, these two
samples differ in stellar mass by $\approx$0.5 dex. If instead we consider
the relative offset in metallicity against the \MZ\ relation of
\citetalias{and13}, the K-S test finds that the two samples are different at
\KSprob\ (1.9$\sigma$; Figure~\ref{fig:7}(b)). The difference, however, is
in the opposite direction of local predictions, with higher sSFR galaxies
having higher gas-phase metallicities.

Given this discrepancy, it is also important to investigate whether the
same result is seen when considering the FMR that \cite{man10} defined.
Here, they utilized the \Pagel\ strong-line diagnostic and a fourth-order
polynomial for \Pagel--$Z$ \citep{mai08}. However, the majority of our
DEEP2 sample have dust-corrected $R_{23}$ values that exceed the maximum
threshold value of $R_{23}=8.65$. Thus, it is not possible
to conduct the same K-S test analysis with \Pagel\ and the \cite{mai08}
calibration.
   
\section{DISCUSSIONS}
\label{sec:dis}

From DEEP2 spectra of \Ndet\ galaxies with oxygen abundances from \OIIIa\
detections (i.e., the \Te\ method), we find that metal-poor strongly
star-forming galaxies are consistent with the local FMR \citepalias{and13},
albeit with large dispersion (0.29 dex with 0.16 dex due to measurement errors).
This result is consistent with metal-poor galaxies from \cite{ly14}, and lensed
low-mass star-forming galaxies at $z\sim$0.8--2.6 \citep{wuy12}. Given the high
sSFRs of $\sim$(100 Myr)$^{-1}$, we argue that the large dispersion in metallicity
is unsurprising---these galaxies are most likely undergoing episodic star
formation and have not settled into a steady state.

We find marginal (\KSprob; 1.9$\sigma$) evidence that galaxies with
higher sSFRs ($\lesssim$10$^{-8}$ yr$^{-1}$) are more metal-rich.
While this contradicts previous local studies, the inverse of the
sSFR---timescale for star formation---is short. Assuming outflow velocities
comparable to virial velocities ($\sim$150 km s$^{-1}$) for
$\log{(M_{\rm halo}/M_{\sun})}\approx11.1$ \citep{beh10}, 8 galaxies in
our sample would not have enough time (sSFR$^{-1}\lesssim10^{7.65}$ yr)
for any recently enriched outflows to be driven out of the 1\arcsec\ (7.5 kpc)
slit-widths. Thus, one would expect the SFR--$Z$ dependence to turn positive
for low-mass strongly star-forming galaxies.
Given the instantaneous SFRs, we find that the measured oxygen abundances
can be explained with low nucleosynthesis yields ($y\sim0.01$),
gas-to-stellar mass fraction of $\approx1\pm0.4$, and no metal loss due to
outflows.

\acknowledgements
Based on observations taken at the W. M. Keck Observatory, which is operated
jointly by the National Aeronautics and Space Administration (NASA), the
University of California, and the California Institute of Technology.
Funding for the DEEP2 Galaxy Redshift Survey has been provided by NSF grants
AST-9509298, AST-0071048, AST-0507428, and AST-0507483 as well as NASA LTSA
grant NNG04GC89G.
C.L. is funded through the NASA Postdoctoral Program. We thank Jeffrey
Newman, Alaina Henry, Massimo Ricotti, and Kate Whitaker for insightful
discussions, and the anonymous referee for insightful comments that
improved the paper.

\clearpage

\newcommand{\ps}{\phm{1}}
\newcommand{\pt}{\phm{10}}
\begin{landscape}
\vspace{1cm}
\begin{deluxetable*}{lccccccccccccc}
  \tabletypesize{\scriptsize}
  \tablewidth{0pc}
  \tablecaption{Summary of DEEP2 Sample}
  \tablehead{
    \colhead{ID}&
    \colhead{R.A.}&
    \colhead{Declination}&
    \colhead{$z$}&
    \colhead{EW(\Hb)}&
    \colhead{$\log(M_{\star}/M_{\sun})$}&
    \colhead{$\log\left[\frac{{\rm SFR}_{\Hbe}}{M_{\sun}~{\rm yr}^{-1}}\right]$}&
    \colhead{\EBVa}&
    \colhead{S/N}&
    \colhead{\OII/\Hb}&
    \colhead{\OIII/\Hb}&
    \colhead{$\frac{\OIII}{\OIII\lambda4363}$}&
    \colhead{$\log(T_e/{\rm K})$}&
    \colhead{\OH}\\
    \colhead{}&
    \colhead{(hh:mm:ss)}&
    \colhead{(dd:mm:ss)}&
    \colhead{}&
    \colhead{(\AA)}&
    \colhead{}&
    \colhead{}&
    \colhead{(mag)}&
    \colhead{($\lambda$4363)}}
  \startdata
01 & 14:18:31.260 & 52:49:42.545 & 0.8194 & \ps38.46 & 8.81$^{+0.18}_{-0.00}$ &           0.23$\pm$0.16 & 0.00$^{+0.09}_{-0.08}$ &    4.9 & 2.165$^{+0.101}_{-0.051}$ & 5.507$^{+0.156}_{-0.067}$ &  57.273$^{+11.748}_{-08.811}$ &          4.18$\pm$0.04 & 7.96$^{+0.11}_{-0.14}$\\[1mm]
02 & 14:21:21.513 & 53:01:07.672 & 0.7496 & \ps51.94 & 8.76$^{+0.04}_{-0.06}$ &  0.54$^{+0.45}_{-0.33}$ & 0.28$^{+0.24}_{-0.18}$ &    5.9 & 1.821$^{+0.705}_{-0.235}$ & 7.792$^{+0.200}_{-0.351}$ &  67.843$^{+19.856}_{-06.619}$ &          4.14$\pm$0.04 & 8.13$\pm$0.10\\[1mm]
03 & 14:21:25.487 & 53:09:48.071 & 0.7099 &   104.49 & 8.94$^{+0.00}_{-0.15}$ & -0.09$^{+0.19}_{-0.25}$ & 0.00$^{+0.10}_{-0.14}$ &    3.7 &           2.413$\pm$0.103 & 5.178$^{+0.198}_{-0.085}$ &  54.985$^{+20.947}_{-10.473}$ & 4.17$^{+0.06}_{-0.05}$ & 7.98$^{+0.17}_{-0.19}$\\[1mm]
04 & 14:22:03.718 & 53:25:47.766 & 0.7878 & \ps 7.38 &                 \ldots & -0.06$^{+0.63}_{-0.59}$ & 0.00$^{+0.34}_{-0.32}$ &    5.5 &           0.867$\pm$0.094 & 7.273$^{+0.385}_{-0.578}$ &  18.192$^{+05.923}_{-01.692}$ & 4.49$^{+0.08}_{-0.07}$ & 7.35$^{+0.11}_{-0.17}$\\[1mm]
05 & 14:21:45.408 & 53:23:52.699 & 0.7710 & \ps74.51 & 8.86$^{+0.07}_{-0.00}$ &  1.50$^{+0.07}_{-0.05}$ & 0.47$^{+0.04}_{-0.03}$ &   10.0 &           1.960$\pm$0.084 & 8.497$^{+0.080}_{-0.053}$ &  90.482$^{+10.167}_{-08.133}$ &          4.10$\pm$0.02 & 8.27$^{+0.07}_{-0.06}$\\[1mm]
06 & 16:47:26.188 & 34:45:12.126 & 0.7166 & \ps36.91 & 8.85$^{+0.12}_{-0.01}$ &  0.71$^{+0.07}_{-0.10}$ & 0.21$^{+0.04}_{-0.05}$ &    5.0 & 2.539$^{+0.221}_{-0.095}$ &           6.245$\pm$0.081 & 131.237$^{+47.723}_{-15.908}$ & 4.02$^{+0.02}_{-0.03}$ & 8.51$^{+0.13}_{-0.10}$\\[1mm]
07 & 16:46:35.420 & 34:50:27.928 & 0.7624 & \ps34.71 & 9.07$^{+0.19}_{-0.00}$ &  0.84$^{+0.12}_{-0.10}$ & 0.22$^{+0.07}_{-0.05}$ &    3.4 & 3.304$^{+0.208}_{-0.260}$ & 5.003$^{+0.040}_{-0.081}$ &  83.493$^{+25.690}_{-12.845}$ &          4.09$\pm$0.05 & 8.29$^{+0.15}_{-0.21}$\\[1mm]
08 & 16:47:26.488 & 34:54:09.770 & 0.7653 & \ps79.33 & 8.07$^{+0.69}_{-0.14}$ &           0.15$\pm$0.42 &          0.22$\pm$0.23 &    4.0 & 1.412$^{+0.779}_{-0.097}$ & 7.866$^{+0.589}_{-0.168}$ &  24.249$^{+08.434}_{-04.217}$ & 4.38$^{+0.08}_{-0.11}$ & 7.63$^{+0.17}_{-0.21}$\\[1mm]
09 & 16:49:51.368 & 34:45:18.210 & 0.7909 & \ps53.06 & 9.00$^{+0.08}_{-0.32}$ &  0.22$^{+0.17}_{-0.12}$ & 0.02$^{+0.09}_{-0.07}$ &    3.6 & 2.496$^{+0.197}_{-0.079}$ & 5.485$^{+0.124}_{-0.099}$ &  78.655$^{+29.964}_{-14.982}$ & 4.10$^{+0.05}_{-0.04}$ & 8.23$^{+0.16}_{-0.17}$\\[1mm]
10 & 16:51:31.472 & 34:53:15.964 & 0.7945 & \ps64.52 & 8.70$^{+0.15}_{-0.31}$ &           1.06$\pm$0.08 &          0.31$\pm$0.04 &    6.4 & 2.159$^{+0.093}_{-0.140}$ & 6.739$^{+0.094}_{-0.040}$ &  85.109$^{+15.474}_{-12.379}$ & 4.11$^{+0.02}_{-0.03}$ & 8.21$^{+0.09}_{-0.08}$\\[1mm]
11 & 16:50:55.342 & 34:53:29.875 & 0.7980 & \ps94.85 & 8.56$^{+0.21}_{-0.59}$ &  0.11$^{+0.16}_{-0.11}$ & 0.00$^{+0.08}_{-0.06}$ &    7.7 &           2.086$\pm$0.065 & 7.480$^{+0.148}_{-0.118}$ &  58.301$^{+03.195}_{-11.181}$ &          4.19$\pm$0.03 & 7.97$^{+0.06}_{-0.11}$\\[1mm]
12 & 16:53:03.486 & 34:58:48.946 & 0.7488 & \ps69.05 & 8.33$^{+0.16}_{-0.08}$ &  0.31$^{+0.10}_{-0.15}$ & 0.06$^{+0.05}_{-0.08}$ &   13.2 & 1.702$^{+0.229}_{-0.000}$ & 6.384$^{+0.130}_{-0.056}$ &  70.872$^{+04.765}_{-05.956}$ & 4.15$^{+0.02}_{-0.01}$ & 8.04$^{+0.03}_{-0.05}$\\[1mm]
13 & 16:51:24.060 & 35:01:38.740 & 0.7936 & \ps32.65 & 9.09$^{+0.09}_{-0.35}$ &  0.40$^{+0.17}_{-0.12}$ & 0.23$^{+0.09}_{-0.06}$ &    5.2 & 2.862$^{+0.369}_{-0.185}$ & 6.241$^{+0.154}_{-0.103}$ &  74.343$^{+15.250}_{-11.437}$ & 4.13$^{+0.03}_{-0.04}$ & 8.20$^{+0.12}_{-0.11}$\\[1mm]
14 & 16:51:20.343 & 35:02:32.628 & 0.7936 &   135.97 & 8.68$^{+0.50}_{-0.30}$ &  0.98$^{+0.08}_{-0.06}$ &          0.21$\pm$0.04 &    9.3 &           2.173$\pm$0.089 &           7.209$\pm$0.040 & 105.147$^{+12.086}_{-09.669}$ & 4.07$^{+0.01}_{-0.02}$ & 8.33$\pm$0.06\\[1mm]
15 & 23:27:20.369 & 00:05:54.762 & 0.7553 &   741.71 & 8.13$^{+0.01}_{-0.21}$ &  1.29$^{+0.09}_{-0.10}$ &          0.48$\pm$0.05 &   13.6 & 1.034$^{+0.096}_{-0.027}$ & 7.428$^{+0.084}_{-0.042}$ &  72.096$^{+04.614}_{-06.921}$ & 4.15$^{+0.01}_{-0.02}$ & 8.00$^{+0.04}_{-0.05}$\\[1mm]
16 & 23:27:43.140 & 00:12:42.832 & 0.7743 & \ps34.15 & 9.23$^{+0.22}_{-0.20}$ &           0.71$\pm$0.42 &          0.22$\pm$0.23 &    3.4 & 1.958$^{+0.945}_{-0.000}$ &           4.320$\pm$0.124 &  46.844$^{+18.738}_{-12.492}$ &          4.21$\pm$0.07 & 7.85$^{+0.20}_{-0.18}$\\[1mm]
17 & 23:27:29.854 & 00:14:20.439 & 0.7637 & \ps40.62 & 8.39$^{+0.37}_{-0.01}$ &           0.78$\pm$0.19 &          0.32$\pm$0.10 &    4.3 & 2.420$^{+0.596}_{-0.074}$ & 6.508$^{+0.147}_{-0.074}$ &  89.583$^{+28.898}_{-17.339}$ &          4.09$\pm$0.04 & 8.29$^{+0.13}_{-0.15}$\\[1mm]
18 & 23:27:07.500 & 00:17:41.503 & 0.7885 & \ps49.03 & 9.34$^{+0.21}_{-0.29}$ &  0.96$^{+0.14}_{-0.17}$ & 0.34$^{+0.08}_{-0.09}$ &    4.4 & 1.794$^{+0.129}_{-0.194}$ &           7.022$\pm$0.113 & 115.500$^{+37.258}_{-14.903}$ &          4.04$\pm$0.04 & 8.38$^{+0.10}_{-0.15}$\\[1mm]
19 & 23:26:55.430 & 00:17:52.919 & 0.8562 & \ps97.37 & 8.99$^{+0.00}_{-0.07}$ &  0.62$^{+0.13}_{-0.08}$ & 0.00$^{+0.07}_{-0.04}$ &    5.2 & 1.636$^{+0.021}_{-0.026}$ & 7.989$^{+0.066}_{-0.132}$ & 128.952$^{+17.993}_{-23.991}$ & 4.04$^{+0.03}_{-0.02}$ & 8.40$^{+0.08}_{-0.13}$\\[1mm]
20 & 23:30:57.949 & 00:03:38.191 & 0.7842 & \ps87.95 & 8.61$^{+0.07}_{-0.28}$ &  0.66$^{+0.18}_{-0.26}$ & 0.26$^{+0.10}_{-0.14}$ &    4.1 & 2.017$^{+0.276}_{-0.221}$ & 5.575$^{+0.132}_{-0.159}$ &  36.499$^{+14.600}_{-05.840}$ & 4.26$^{+0.05}_{-0.09}$ & 7.72$^{+0.14}_{-0.19}$\\[1mm]
21 & 23:31:50.728 & 00:09:39.393 & 0.8225 & \ps63.83 & 8.81$^{+0.06}_{-0.52}$ &  0.34$^{+0.29}_{-0.30}$ & 0.09$^{+0.16}_{-0.17}$ &    5.0 & 1.692$^{+0.344}_{-0.115}$ & 6.989$^{+0.179}_{-0.223}$ &  52.462$^{+11.343}_{-08.507}$ & 4.19$^{+0.05}_{-0.04}$ & 7.93$^{+0.10}_{-0.14}$\\[1mm]
22 & 02:27:48.871 & 00:24:40.077 & 0.7838 &   298.74 & 7.85$^{+0.45}_{-0.06}$ &  0.30$^{+0.16}_{-0.23}$ & 0.07$^{+0.09}_{-0.13}$ &    5.3 & 1.735$^{+0.343}_{-0.043}$ & 7.110$^{+0.103}_{-0.207}$ &  61.688$^{+12.654}_{-09.490}$ & 4.16$^{+0.03}_{-0.04}$ & 8.04$^{+0.10}_{-0.12}$\\[1mm]
23 & 02:27:05.706 & 00:25:21.865 & 0.7661 & \ps77.52 & 8.63$^{+0.07}_{-0.18}$ &           0.63$\pm$0.42 &          0.22$\pm$0.23 &    3.2 & 1.492$^{+0.720}_{-0.000}$ & 7.631$^{+0.135}_{-0.108}$ & 144.017$^{+00.000}_{-57.607}$ & 4.05$^{+0.05}_{-0.06}$ & 8.41$^{+0.12}_{-0.20}$\\[1mm]
24 & 02:27:30.457 & 00:31:06.391 & 0.7214 &   159.58 & 7.87$^{+0.52}_{-0.29}$ &  0.90$^{+0.07}_{-0.06}$ &          0.25$\pm$0.04 &    9.6 &           1.525$\pm$0.076 &           7.105$\pm$0.057 &  94.916$^{+08.344}_{-10.430}$ &          4.09$\pm$0.02 & 8.20$^{+0.05}_{-0.06}$\\[1mm]
25 & 02:26:03.707 & 00:36:22.460 & 0.7888 & \ps66.11 & 8.69$^{+0.22}_{-0.48}$ &  2.12$^{+0.20}_{-0.17}$ & 1.02$^{+0.11}_{-0.09}$ &    6.6 & 2.857$^{+0.269}_{-0.336}$ & 4.878$^{+0.130}_{-0.065}$ &  48.598$^{+05.116}_{-08.526}$ & 4.23$^{+0.04}_{-0.03}$ & 7.84$^{+0.10}_{-0.09}$\\[1mm]
26 & 02:26:21.479 & 00:48:06.813 & 0.7743 & \ps36.94 & 9.24$^{+0.08}_{-0.29}$ &           0.54$\pm$0.10 & 0.00$^{+0.05}_{-0.06}$ &    6.7 & 2.064$^{+0.031}_{-0.041}$ & 5.895$^{+0.089}_{-0.044}$ &            109.091$\pm$14.307 &          4.07$\pm$0.02 & 8.29$^{+0.08}_{-0.10}$\\[1mm]
27 & 02:29:33.654 & 00:26:08.023 & 0.7294 & \ps67.55 & 8.66$^{+0.20}_{-0.67}$ &  0.80$^{+0.14}_{-0.20}$ & 0.36$^{+0.08}_{-0.11}$ &    7.0 & 2.555$^{+0.220}_{-0.330}$ & 7.849$^{+0.132}_{-0.231}$ &  54.969$^{+09.644}_{-07.715}$ &          4.19$\pm$0.04 & 8.01$^{+0.09}_{-0.12}$\\[1mm]
28 & 02:29:02.031 & 00:30:08.127 & 0.7315 & \ps99.97 & 8.70$^{+0.01}_{-0.06}$ &  1.18$^{+0.03}_{-0.04}$ &          0.30$\pm$0.02 &   15.2 & 1.515$^{+0.042}_{-0.028}$ & 7.633$^{+0.037}_{-0.019}$ & 132.444$^{+10.667}_{-05.333}$ &          4.04$\pm$0.01 & 8.40$^{+0.04}_{-0.03}$\\[1mm]
  \vspace{-3mm}
  \enddata
  \label{tab:1}
  \tablecomments{Unless otherwise specified, \OIII\ refers to \OIII$\lambda\lambda$4959,5007.}
\end{deluxetable*}
\clearpage
\end{landscape}

\end{document}